\documentclass[reprint,aps,showpacs]{revtex4-1}
\usepackage{graphicx,subfig}
\usepackage{mathrsfs}
\usepackage{graphicx}
\usepackage{ragged2e}
\usepackage[font=small]{caption}
\usepackage{caption}
\usepackage{color}
\usepackage{amsmath, amsfonts, amssymb, bm}
\usepackage{braket,cancel}
\begin{document}
	%%%%%%%%%%%%%%%%%%%%%%%%%%%%%%%%%%%%%%%%%%%%%%%%%%%%%%%%%%%%%%%%%
	\title{Strong-field Breit-Wheeler pair production with bremsstrahlung gamma-rays in the perturbative-to-nonperturbative transition regime}
	\author{A. Eckey}
	\author{A.B. Voitkiv}
	\author{C. M\"uller}
	\affiliation{Institut f\"ur Theoretische Physik I, Heinrich Heine Universit\"at D\"usseldorf, Universit\"atsstr. 1, 40225 D\"usseldorf, Germany}
	\date{\today}
	%%%%%%%%%%%%%%%%%%%%%%%%%%%%%%%%%%%%%%%%%%%%%%%%%%%%%%%%%%%%%%%%%
	\begin{abstract}
		Photoproduction of electron-positron pairs by bremsstrahlung gamma-rays propagating through a high-intensity optical laser wave is studied. By calculating total pair production yields as well as energy and angular distributions of the produced particles for a range of laser intensities, we analyze the influence that the broad frequency spectrum of bremsstrahlung exerts on the properties of the process in the transition region between the perturbative and non-perturbative interaction regimes. In particular, we demonstrate a continuous shifting of the mainly contributing interval of bremsstrahlung frequencies during the course of the transition and indicate characteristic signatures of the laser-dressed particle mass.
	\end{abstract}
	%%%%%%%%%%%%%%%%%%%%%%%%%%%%%%%%%%%%%%%%%%%%%%%%%%%%%%%%%%%%%%%%%
	\maketitle
	%%%%%%%%%%%%%%%%%%%%%%%%%%%%%%%%%%%%%%%%%%%%%%%%%%%%%%%%%%%%%%%%%
	\section{Introduction}
	The possibility of converting pure electromagnetic energy into matter in photon-photon collisions represents a clear manifestation of the nontrivial structure of the quantum vacuum.
	In a seminal paper in 1934, Breit and Wheeler described
	the production of an electron-positron pair as the result
	of a collision between two light quanta \cite{BW}. In the 1960s
	the process was generalized to the multiphoton case,
	considering pair production by a light quantum propagating
	through a plane-wave laser field according to the reaction
	$\omega' + n\omega \to e^+e^-$, with the number $n$ of absorbed
	laser photons \cite{Reiss, Ritus, Ritus2, Ritus3}. Various parameter regimes of this strong-field (or nonlinear) version of the Breit-Wheeler process were identified where the pair production exhibits qualitatively different properties. In particular, a distinction between the perturbative and non-perturbative regimes of interaction with the laser field can be
	made, which are mainly characterized by the dimensionless intensity
	parameter $\xi = -e\mathscr{E}_0/(\omega m)$, with the electron charge $e < 0$, its mass $m$, the laser frequency $\omega$ and electric field amplitude $\mathscr{E}_0$. In the perturbative regime with $\xi\ll1$, the pair production rate follows a simple power law of the form $~ \xi^{2n}$.\\
	\\
	A concrete setup to experimentally detect strong-field
	Breit-Wheeler pair creation was proposed in \cite{Reiss2} based on
	the combination of high-energy bremsstrahlung with a few
	intense laser photons. It was oriented towards the properties
	of the highly relativistic electron beam at the Stanford Linear
	Accelerator Center (SLAC). In fact, 25 years later, a very similar experiment was successfully performed at SLAC in the $\xi <1$ range, using electrons close to 50 GeV energy and their Compton scattering from a laser beam of about 2.4 eV frequency \cite{Burke}. In other parameter regimes, however, the process has not been validated experimentally yet.\\
	\\
	At present, there is significant interest in the non-
	perturbative regime, where the production rate for
	$\xi \gtrsim 1$ first starts to deviate from a power-law, and
	for $\xi\gg 1$ turns into an exponential behaviour similar
	to the well-known Schwinger rate \cite{Ritus4, Piazza}. For its realization, the combination of bremstrahlung gamma rays with a
	counterpropagating high-intensity optical laser wave is considered as a feasible experimental pathway. Strong-field Breit-Wheeler pair production in this setup has recently been studied theoretically \cite{Blackburn, Ringwald, Golub} and a
	dedicated experiment has been proposed by the LUXE consortium at DESY \cite{LUXE}. It is going to exploit the high-energy
	electron beam, which usually drives the European X-ray Free-Electron Laser, to generate GeV photons via bremstrahlung from a tungsten target. A particular goal is to determine the transition from the perturbative regime of the process at $\xi \ll 1$ to the non-perturbative regime at $\xi > 1$ (or even $\xi \gg  1$). Experimental campaigns on strong-field QED phenomena are also planned at Rutherford Appleton Laboratory \cite{RAL}, SLAC \cite{E320}, the Center of Advanced Laser Applications \cite{Salgado}, and the Extreme Light Infrastructure \cite{ELI}. \\
	\\
	We note that a number of theoretical studies on strong-field
	Breit-Wheeler pair creation have been carried out in recent years
	using monoenergetic gamma photons and short few-cycle laser pulses
	(see, e.g., \cite{Heinzl,Titov,Krajewska,Meuren,Jansen,BW2}). In this case, the laser field covers a relatively narrow coherent spectrum of frequencies which influences the process properties in a characteristic way. Pair production studies with bremsstrahlung are complementary to this, since here the beam of gamma photons comprises a very broad incoherent spectrum of frequencies. \\
	\\
	In this paper  we study strong-field Breit-Wheeler pair production by bremsstrahlung gamma-photons propagating through a high-intensity
	optical laser field. Our focus lies on the transition from the
	perturbative to the nonperturbative interaction regime around $\xi\approx 1$ and on the impact that the broad frequency spectrum of bremsstrahlung exerts on the process. In this scope, we calculate both the total production yield as well as the energy and angular distributions of the produced positron for a sequence of $\xi$ values. Accordingly, our study differs in particular from
	\cite{Blackburn} that considers the regime of large $\xi\gg1$, and from \cite{Ringwald} that exclusively investigates total production rates.\\
	\\
	Unless otherwise stated, we use relativistic units where $c= \hbar =4\pi\epsilon_0= 1$ and we employ the metric tensor $\text{diag}(g^{\mu \nu})=(1,-1,-1,-1)$. According to this, the four-product of two four-vectors $a^\mu = (a_0, \textbf{a})$ and $b^\mu = (b_0, \textbf{b})$ yields $a\cdot b = a_0 b_0 - \textbf{a}\cdot \textbf{b}$. Four-products with Dirac $\gamma$-matrices are denoted by the Feynman slash notation.
	\section{Theoretical consideration}
	In this section we briefly outline our approach to electron-positron pair production by bremsstrahlung gamma-photons colliding with an intense laser wave, and highlight some particular relations which will be relevant for the subsequent discussions. Our study is motivated by the planned LUXE experiment, where laser pulses of $\lambda = 800$nm wave length and $30$ fs pulse duration are going to be applied \cite{LUXE}. Since these pulses---comprising about 11 field cycles---are not extremely short but rather relatively long, we shall provide and compare predictions from two different field models: the laser field is either treated as a monochromatic wave of infinite extent (see Sec.II.A) or as a finite pulse with slowly varying envelope (see Sec.II.B).
	\subsection[A.]{Infinitely extended, monochromatic laser wave}
	The S-matrix element for the strong-field Breit-Wheeler  process with a high-energy photon of frequency $\omega^\prime$, wave vector ${k^\prime}^\mu$ and polarization vector ${\varepsilon^\prime}^\mu$ is given by \cite{Ritus3,Greiner}
	\begin{eqnarray}
		S_{\text{fi}} = -ie\sqrt{\frac{4\pi}{2\omega^{\prime}V}}\int \text{d}^4x \text{e}^{-ik^{\prime} \cdot x} \bar{\Psi}_{ \textbf{p}^{\prime},\text{s}^\prime} \cancel{\varepsilon}^\prime \Psi_{-\textbf{p},\text{s}} 
	\end{eqnarray}
	For the initial and final state we apply the Volkov wave functions
	\begin{eqnarray}
		& &\Psi_{\textbf{p}^\prime,s^\prime}(x)=\sqrt{\frac{m}{q_0 V}} \left(1+\frac{e}{2k\cdot p^\prime}\cancel{k}\cancel{A}\right)u(p^\prime,s^\prime)  \\
		& &\times \exp(-iea\frac{\varepsilon_1 \cdot p^\prime}{k \cdot p^\prime}\sin(k\cdot x)+ iea\frac{\varepsilon_2 \cdot p^\prime}{k \cdot p^\prime}\cos(k\cdot x)-iq^\prime\cdot x) \notag
	\end{eqnarray}
	with the electron momentum ${p^\prime}^\mu=(E_{p^\prime},\textbf{p}^\prime)$ and the normalization volume $V$. The corresponding positronic Volkov state can be obtained by a replacement of ${p^\prime}^\mu$ with $-p^\mu$ where $\textbf{p}$ is referred to as the positron momentum. The effective momentum $q^\mu =p^\mu + \xi^2 m^2/(2k\cdot p)k^\mu$ fullfills the dispersion relation $q^2=m_\star^2$ where $m_\star=m\sqrt{1+\xi^2}$ denotes the laser-dressed mass. We consider the laser field here as an infinitely extended, monochromatic wave of circular polarization with the four-potential
	\begin{eqnarray}
		A^\mu(x)=a(\varepsilon_1^\mu \cos k\cdot x +\varepsilon_2^\mu \sin k\cdot x) \notag
	\end{eqnarray}
	in the Lorenz gauge $k \cdot A=0$. As will be shown below the monochromatic field approach is well applicable to the laser parameters of interest here. The corresponding wave vector of the counter-propagating laser photons and the two corresponding transverse and orthogonal polarization vectors are taken as $k^\mu=\omega(1,0,0,1)$, $\varepsilon^\mu_1=(0,1,0,0)$ and $\varepsilon^\mu_2 = (0,0,1,0)$. Using this, the production rate per $\gamma$-photon is obtained as 
	\begin{eqnarray}
		&&\text{d}R(k,k^\prime)= \frac{|S_{\text{fi}}|^2}{T}V\frac{\text{d}^3q}{(2\pi)^3}V\frac{\text{d}^3q^\prime}{(2\pi)^3}\\
		&=&\frac{e^2}{2\pi}\frac{m^2}{q_0 q^{\prime}_0 \omega^{\prime}}\sum_{n=n_0}^{\infty} \delta^4(q+q^{\prime}-k^{\prime}-nk) B_n(q^\prime,q,\xi)\text{d}^3q \text{d}^3q^{\prime} \notag 
	\end{eqnarray}
	where the sum over $n$ results from a Fourier series expansion of the periodic parts in the S-matrix element. Here
	\begin{eqnarray}
		&&B_n(q^\prime,q,\xi) = \text{J}_n^2(z)\\
		&&+\xi^2\left(\text{J}_n^2(z)-\frac{1}{2}\text{J}_{n-1}^2(z)-\frac{1}{2}\text{J}_{n+1}^2(z)\right)\left(1-\frac{(k\cdot k^\prime)^2}{2k\cdot pk\cdot p^\prime}\right)\notag
	\end{eqnarray}
	with $z=ea\sqrt{-[q/(k\cdot q)-q^\prime/(k\cdot q^\prime)]^2}$ and $n_0$ denotes the minimal necessary photon number which depends on the laser-dressed mass by virtue of
	\begin{eqnarray}
		n_0 \geq \frac{2 m_\star^2}{k \cdot k^\prime}.
	\end{eqnarray}
	For the special case of a head-on collision, that we will consider in the following, one has $k\cdot k^\prime = 2\omega \omega^\prime$.\\
	\\
	Equation (3) represents the unpolarized differential rate, since we have averaged over the photon polarization and have summed over the lepton spins $s$ and $s^\prime$. The total rate is obtained by a sixfold momentum integration. We will call $R(k,k^\prime)$ the 'monochromatic' rate as it refers to a fixed $\gamma$-photon frequency.  Using the four $\delta$-functions one can also formulate the differential production rates
	\begin{eqnarray}
		\frac{\text{d}R}{\text{d}\vartheta_{q}}&=&\sum_{i=\pm} \sum_{n=n_0}^{\infty} {Q_i}^2\frac{e^2 m^2}{q_0 q_0^\prime \omega^\prime}\notag \\
		& & \times \frac{B_n(q^\prime,q,\xi)|_{|\textbf{q}|=Q_i}}{\left|\frac{Q_i}{\sqrt{{Q_i}^2+m_{\star}^2}}+ \frac{Q_i+p_n \cos\vartheta_q}{\sqrt{p_n^2+{Q_i}^2+2p_n Q_i \cos\vartheta_q+m_{\star}^2}} \right|} \notag 
	\end{eqnarray}
	with respect to the polar angle and 
	\begin{eqnarray}
		\frac{\text{d}R}{\text{d}E_{q}}&=& \frac{e^2 m^2}{\omega^\prime} \sum_{n=n_0}^{\infty} \frac{1}{\omega^\prime-n\omega} B_n(q^\prime,q,\xi)|_{\cos\vartheta_q=\cos\vartheta_0} \notag
	\end{eqnarray}
	with respect to the energy of the created positron, where
	\begin{eqnarray}
		\frac{Q_\pm}{m_\star}&=& \frac{(\rho_n^2-1)\rho_n\chi_n\cos\vartheta_q}{1-\rho_n^2\cos^2\vartheta_q} \notag \\
		&\pm& \frac{\sqrt{\chi_n^2(1-\rho_n^2)^2-1+(\rho_n \cos\vartheta_q)^2}}{1-\rho_n^2\cos^2\vartheta_q}
	\end{eqnarray}
	and 
	\begin{eqnarray}
		\cos(\vartheta_0)=\frac{2n\omega\omega^\prime-E_n E_{q}}{|\textbf{q}|(\omega^\prime-n\omega)}.
	\end{eqnarray}
	To simplify the notation we have introduced the quantities $p_n=\omega^\prime-n\omega$, $E_n=\omega^\prime+n\omega$, $\rho_n=p_n/E_n$ and $\chi_n=E_n/2m_\star$. With the help of Eqs. (6) and (7) it is possible to determine, for given $n$, the minimal polar emission angle and the accessible range of positron energies. Since the argument of the square root should be positive and $|\cos(\vartheta_0)|\le 1$, these conditions can be stated as
	\begin{eqnarray}
		\vartheta_{\text{min}}= \pi - \arccos \left(\sqrt{\frac{1}{\rho_n^2}-\frac{\chi_n^2}{\rho_n^2}(1-\rho_n^2)^2}\right)
	\end{eqnarray}
	and
	\begin{eqnarray}
		E_{\text{min}}=\frac{E_n}{2}-\Omega\leq E_{q\prime} \leq \frac{E_n}{2}+\Omega = E_{\text{max}}
	\end{eqnarray}
	with
	\begin{eqnarray}
		\Omega = \frac{p_n}{2}\sqrt{\frac{n\omega\omega^\prime-m_\star^2}{n\omega\omega^\prime}}.
	\end{eqnarray}
	\subsection[B.]{Finite plane-wave laser pulse}
	In the previous subsection we outlined the theory for nonlinear Breit-Wheeler pair creation in an infinitely extended, monochromatic plane laser wave. In experimental reality, laser pulses have a finite extent that can be incorporated in the theoretical treatment directly on the basis of the Volkov states \cite{Heinzl,Titov,Krajewska,Meuren,Jansen,BW2}. For relatively long laser pulses, as considered here, one may follow an alternative approach which essentially relies on an average of monochromatic pair production rates over the envelope of a finite laser pulse. This method has recently been derived thoroughly and termed 'locally monochromatic approximation' (LMA) \cite{King}. It is applicable when the pulse envelope varies very slowly on the time scale of the field oscillations \cite{Bamber}.\\
	\\
	Let the circularly-polarized plane-wave laser pulse be described by a four-potential
	$$ A^\mu(x) = a f(\phi / \Phi)(\varepsilon_1^\mu \cos k\cdot x +\varepsilon_2^\mu \sin k\cdot x) $$
	where now $\phi=k\cdot x$ is the laser phase, $\pi \Phi$ the phase duration of the pulse, and $f(\phi / \Phi)$ the envelope function. The latter is assumed to have compact support on the interval $-N\pi\le\phi\le N\pi$ so that $\Phi=2 N$, with the number $N$ of field oscillation cycles. Within the LMA, the probability for nonlinear Breit-Wheeler pair production in a laser pulse of this form reads \cite{King}
	\begin{eqnarray}
		W(k,k') &=& \frac{e^2 m^2}{k \cdot k^\prime} \int d\phi \sum_{n=n_0(\phi)}^\infty  \notag \\
		&  &\times \int_{r_-(\phi)}^{r_+(\phi)}dr B_n(q^\prime,q,f(\phi/\Phi)\xi) 
	\end{eqnarray}
	with $r=(k\cdot p)/(k\cdot k^\prime)$,
	\begin{eqnarray}
		z(\phi) &&= \frac{2 n \xi |f(\phi/\Phi)|}{\sqrt{1+\xi^2f^2(\phi/\Phi)}} \notag \\
		&&\bigg[\frac{\big(1+\xi^2 f^2(\phi/\Phi)\big)m^2}{2 n (k\cdot k^\prime) (1-r)r} \bigg(1 - \frac{\big(1+\xi^2 f^2(\phi/\Phi)\big)m^2}{2 n (k\cdot k^\prime) (1-r)r} \bigg)\bigg]^{1/2} \notag
	\end{eqnarray}
	and the integration boundaries $r_\pm = 1/2 [1\pm\sqrt{1- n_0(\phi)/n}]$. The minimally required laser photon number [see Eq.(5)] now depends on $\phi$ since the laser intensity (slowly) varies over the pulse envelope ($n_0(\phi) \geq \frac{2 m^2(1+\xi^2 f^2(\phi/\Phi))}{k \cdot k^\prime}$). Also the effective momenta $q^\mu$, $q'^\mu$ and the dressed mass $m_\star$ become $\phi$-dependent via the replacement $\xi^2 \to \xi^2 f^2(\phi/\Phi)$ \cite{Harvey}.  Besides, $\omega$ represents here the central frequency of the pulse. Since the latter has finite extent it contains a range of frequencies which, however, is rather narrow for long pulse durations. For the same reasons, relations like those for the minimum emission angle $\vartheta_\text{min}$ in Eq.(8), that strictly hold for infinitely extended monochromatic laser waves, are smeared out when the pair production occurs in a finite laser pulse. Nevertheless, in the case of long pulse durations, they remain physically meaningful, as we will show in Sec.~III.\\
	It is worth pointing out that the laser field models in Secs.II.A and II.B disregard focussing effects, which lead to deviations from a plane-wave form. These deviations can be measured by the beam divergence, given by the ratio between the focal waist size $w_0$ and the Rayleigh length $z_R=\pi w_0^2/\lambda$. The intermediate regime of $\xi\sim 1$ will be probed in 'phase 0' at LUXE with a laser beam of 40\,TW power \cite{LUXE}. A rather moderate focussing down to $w_0=8\,\mu$m already leads to $\xi\approx 3$ which is the highest value considered in our study. The beam divergence then is $w_0/z_R\approx 0.03$ and thus very small. For smaller values of $\xi$, the associated beam divergence will be even smaller. As a consequence, laser focussing effects may be neglected in our treatment.
	\subsection[C.]{Inclusion of Bremsstrahlung}
	Since the incoming $\gamma$-photon is produced via bremsstrahlung we want to obtain the pair creation rate resulting from a collision of this bremsstrahlung photons with a laser photon. Therefore we integrate the rate, as given in Eq. (3), weighted by the distribution function of the bremsstrahlung photons with respect to the photon momentum. Thus, the averaged rate is given by
	\begin{eqnarray}
		\bar{R}(k)=\int \frac{\text{d}^3k^\prime}{(2\pi)^3}w_\gamma(k^\prime)R(k,k^\prime)
	\end{eqnarray}
	with the bremsstrahlung distribution function in spherical coordinates
	\begin{eqnarray}
		w_\gamma(k^\prime)=\frac{(2\pi)^3 I_\gamma(f,\ell)}{{\omega^\prime}^2\sin(\theta_{k^\prime})E_0}\Theta(E_0-\omega^\prime)\delta(\theta_{k^\prime}-\vartheta_{k^\prime})\delta(\varphi_{k^\prime}) \notag
	\end{eqnarray}
	where $E_0$ represents the initial kinetic energy of the incoming electrons, which are used for the production of the bremsstrahlungs photons, $\theta_{k^\prime}$ labels the angle of propagation of the bremsstrahlungs photons and $\Theta$ denotes the Heaviside function. $I_\gamma(f,\ell)$ is the photon frequency spectrum derived within the complete screening approximation \cite{Blackburn}
	\begin{eqnarray}
		I_\gamma(f,\ell)\approx \frac{(1-f)^{\frac{4}{3}\ell}-\text{e}^{-\frac{7}{9}\ell}}{f(\frac{7}{9}+\frac{4}{3}\ln(1-f))}. \notag
	\end{eqnarray}
	The photon energy spectrum depends on the normalized target thickness $\ell=L_T/L_{\text{rad}}$ with respect to the radiation length of the target material $L_{\text{rad}}$. It also includes the normalized photon energy $f=\omega^\prime/E_0$. For the ranges $f\geq 0.2 = f_{\text{min}}$ and $0.5\leq \ell \leq2$, the formula given above turns out to be a good estimation. The considered incoming electrons are highly relativistic. For this reason the incident electron propagation direction is the preferred directon of the emitted bremsstrahlung photons. In our case of a head-on collision one has $\theta_{k^\prime}=\pi$. Taking into account all these assumptions, the rate of the process is given by
	\begin{eqnarray}
		\bar{R}(k)=\int_{f_{\text{min}}}^{1}\text{d}f R(k,k^\prime) I_\gamma(f,\ell).
	\end{eqnarray}
	To obtain the number of pairs created per incident electron in experiment, the rate (13) is multiplied by the interaction time, the focal volume and the gamma photon density. The latter depends on the bremsstrahlung beam divergence and the distance of the converter target to the interaction point \cite{Golub}. In the case of a finite laser pulse, a relation analogous to Eq.(13) applies to the production probability averaged over the bremsstrahlung, $\bar W(k)$, with the rate $R(k,k')$ being replaced by $W(k,k')$ from Eq.(11).
	\section{Results and Discussion}
	We have calculated the strong-field Breit-Wheeler pair production
	with bremsstrahlung gamma photons for laser intensity parameters
	in the range $0.2 \le \xi \le 3.0$. Two different scenarios in terms of the involved photon frequencies are considered. In the first scenario, taking the parameters of the planned LUXE experiment as reference \cite{LUXE}, we set the energy of the incident electron beam generating bremsstrahlung to $E_0=16.5$\,GeV and the laser photon frequency to $\omega = 1.5$\,eV. Besides, in order to study the characteristics of the pair production process at considerably higher frequencies, we consider $E_0=70$\,GeV and $\omega=2.4$\,eV in a second scenario. Note, for comparison, that electrons with energies up to 100\,GeV are nowadays available at CERN \cite{X,Y}. A normalized target thickness of $\ell=0.5$ is assumed throughout.\\
	Total pair production rates averaged over the bremsstrahlung spectrum are shown in Tab.\,I for various $\xi$ values. In the first scenario, the rate grows very steeply with increasing $\xi$. This is because the relevant field strength in the boosted frame $\mathcal{E}'\sim (E_0/m)\mathcal{E}_0$ always remains undercritical, reaching 30\% of the Schwinger field $\mathcal{E}_{\rm S} = m^2/|e|$ at $\xi=3$. Conversely, for the chosen parameters in the second scenario, the pair production occurs around $\mathcal{E}'\sim \mathcal{E}_{\rm S}$, leading to vastly larger rates and a much slower rate growth.
	\begin{table}[h]
		\centering
		\begin{tabular}{ |p{0.5cm}||p{1.3cm}|p{1.3cm}||p{1.3cm}|p{1.3cm}|}
			\hline
			\parbox[c]{1cm}{ ~ \newline  ~ \newline  ~ \newline} & \multicolumn{2}{|c||}{\parbox[c]{2.5cm}{ $E_0=16.5$\,GeV \newline $\omega=1.5$\,eV}} & 
			\multicolumn{2}{|c|}{\parbox[c]{2.5cm}{$E_0=70.0$\,GeV \newline $\omega=2.4$\,eV} } \\
			\hline
			$\xi$ & $\bar{R}$\,  $[\text{s}^{-1}]$& $\omega^\prime_{\rm eq}$\,$[\text{GeV}]$&$\bar{R}$\,$[\text{s}^{-1}]$&$\omega^\prime_{\rm eq}$\,$[\text{GeV}]$ \\
			\hline
			0.2  &2.5$\times10^{-9}$&14.6&2.2$\times10^{8}$ &44.4\\
			0.4  &1.5$\times10^{-2}$&13.9&3.8$\times10^{9}$ &39.6\\
			0.7  &6.4$\times10^{2}$ &13.1&3.9$\times10^{10}$&35.2\\
			1.0  &1.5$\times10^{5}$ &12.5&1.4$\times10^{11}$&31.7\\
			1.5  &2.2$\times10^{7}$ &11.6&5.2$\times10^{11}$&27.8\\
			3.0  &7.7$\times10^{9}$&9.8&3.2$\times10^{12}$&21.5 \\
			\hline
		\end{tabular}
		\caption{Total rate $\bar{R}$ averaged over the bremsstrahlung spectrum for different intensity parameters $\xi$ with laser photon energy $\omega$ and incident electron energy $E_0$ to generate the bremsstrahlung photons. Besides, $\omega_{\rm eq}^\prime$ gives the gamma-photon energy at which the corresponding monochromatic rate R equals the averaged rate $\bar{R}$. }
	\end{table}
	It is an interesting question at which equivalent gamma-photon frequency $\omega'_{\rm eq}$ the associated {\it monochromatic} production rate $R(k,k')$ equals the corresponding bremsstrahlung-averaged rate $\bar{R}(k)$. It turns out that these equivalent frequencies decrease substantially when $\xi$ increases (see Tab.\,I). Qualitatively this phenomenon can be understood by noting that a larger $\xi$ value facilitates the absorption of a larger number of laser photons, so that lower gamma-photon energies become sufficient to create pairs.\\
	\begin{figure}[h]
		\centering
		\subfloat[][]{\includegraphics[width=0.9\linewidth]{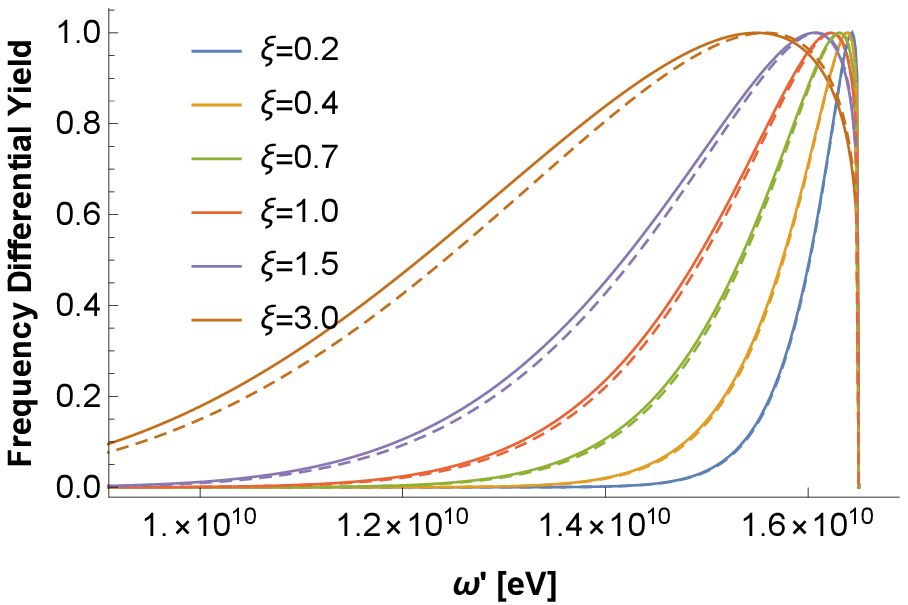}}
		\qquad
		\subfloat[][]{\includegraphics[width=0.9\linewidth]{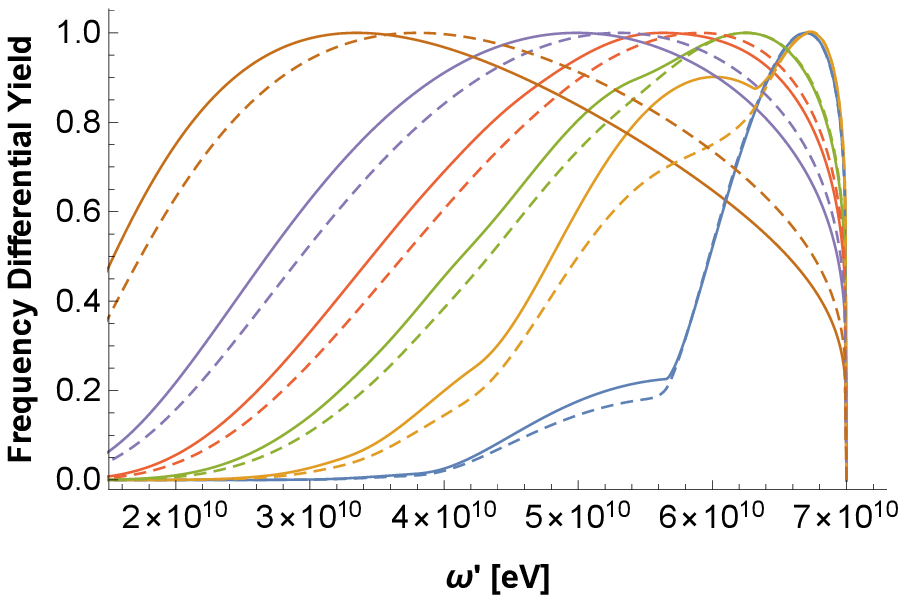}}
		\caption{Solid graphs show $\text{d}\bar{R}/\text{d}\omega^\prime$ for an infinite laser wave, whereas dashed graphs display $\text{d}\bar{W}/\text{d}\omega^\prime$ for an 11-cycle, cos$^2$-shaped laser pulse, as a function of the energy of the $\gamma$-photon for $\xi=0.2,0.4,0.7,1.0,1.5$ and $3.0$ (see color coding). Panel (a) shows the results for $E_0=16.5$GeV and $\omega=1.5$eV whereas panel (b) for $E_0=70.0$GeV and $\omega=2.4$eV. The curves have been normalized to $1$ in order to facilitate a comparison of their shapes.}
	\end{figure}
	\\
	Deeper insights can be gained by inspection of the underlying differential rates, which are discussed next. We point out that all results in the following graphs have been normalized to $1.0$. This way all curves can be shown on common scales to facilitate their comparison.\\
	\\
	Figure 1 shows the differential rate $d\bar{R}/d\omega^\prime$ as a
	function of the energy of the bremsstrahlungs photon $\omega^\prime$. We see that as the intensity increases, the largest contributions to the rate stem from bremsstrahlung photons that tend to carry smaller energies, which is consistent
	with the behaviour of $\omega'_{\rm eq}$ in Tab.~I. In the considered range of parameters, the monochromatic rate $R(k,k^\prime)$ grows with increasing value of $\omega^\prime$, whereas the bremsstrahlung spectrum falls when the energy rises.
	This interplay determines the respective $\omega^\prime$ range that gives the main contribution to the total averaged rate.
	When $\xi\ll 1$, the dominant contribution stems from bremsstrahlung energies close to $E_0$ because this way
	the required number of laser photons is minimized. When $\xi$ is enlarged, the $d\bar{R}/d\omega^\prime$ curves become broader and shift towards smaller energies. This is a clear indication of the transition to the nonperturbative regime, where a broad range of laser photon numbers $n$, lying considerably above $n_0$, give relevant contributions to the pair production \cite{Mueller}. To quantify the shifting behavior further, one can calculate
	the average gamma-photon energy that contributes to the rate
	$\bar R$, according to $\langle \omega' \rangle = \frac{1}{\bar R}
	\int \frac{d\bar R}{d\omega'}\, \omega'\,d\omega'$. It attains
	the values $\langle \omega' \rangle = 16.1$, 15.2 and 13.7\,GeV
	for $\xi = 0.2$, 1.0 and 3.0, respectively.\\
	\\
	Moreover, in Fig.~1\,(b) for $E_0 = 70$\,GeV, a stepwise behavior is observed in the few-photon regime for $\xi < 1$ that can also be explained by the mainly contributing laser photon numbers. For example, the minimum required photon number changes from 3 to 2 photons at $\omega^\prime \approx 57$\,GeV and $\xi=0.2$ and from 4 to 3 photons at $\omega^\prime \approx 38$\,GeV. At $\xi=0.4$, the change from 3 to 2 photons is seen correspondingly
	at $\omega^\prime \approx 63$\,GeV. In the course of the transition to the nonperturbative regime with $\xi\gtrsim 1$, the pronounced steps are washed out and disappear. Note that in Fig.~1\,(a) for the LUXE parameters, a corresponding steplike behavior is not visible because the number of participating photons is much higher there.
	\\
	In addition to the rates, which are based on the calculations with an infinitely extended laser wave, the differential probabilities $d\bar{W}/d\omega^\prime$, calculated in the LMA, are shown. The envelope function is given by $f(\phi/\Phi)=\cos^2(\phi/\Phi)$ and contains N=11 field cycles.
	For the frequency-differential yields, the inclusion of the finite laser pulse duration causes only minor modifications. While for $E_0=70$ GeV the shape of the curves remains essentially unaltered, small overall shifts towards higher frequencies $\omega^\prime$ occur for $\xi\gtrsim 1$, and solely for $\xi=0.4$ some more pronounced differences appear in the frequency range $\omega^\prime \approx 50 -65$GeV. Instead, in case of the LUXE parameters, the LMA curves very closely agree with those from the infinitely extended, monochromatic laser wave.
	\\
	\begin{figure}[]
		\centering
		\subfloat[][]{\includegraphics[width=0.9\linewidth]{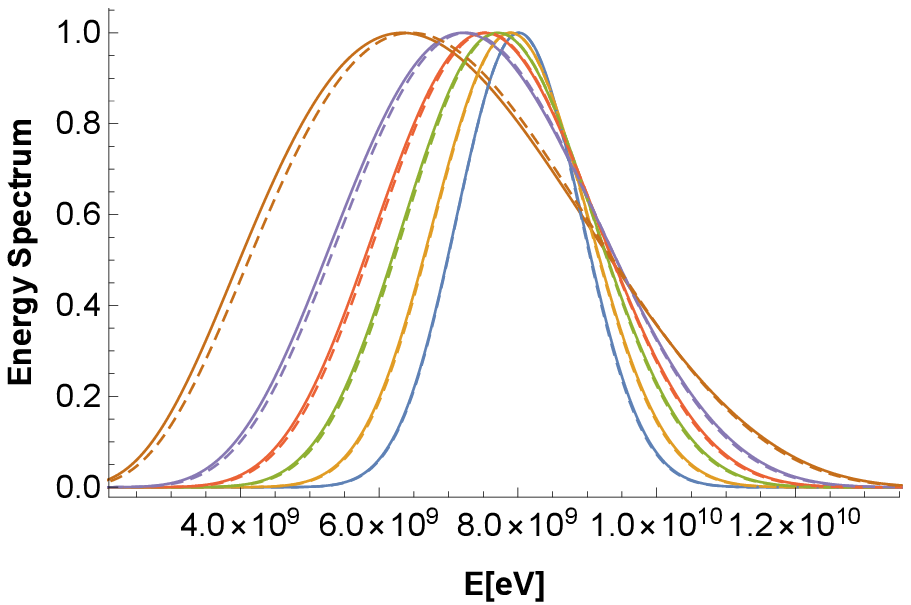}}
		\qquad
		\subfloat[][]{\includegraphics[width=0.9\linewidth]{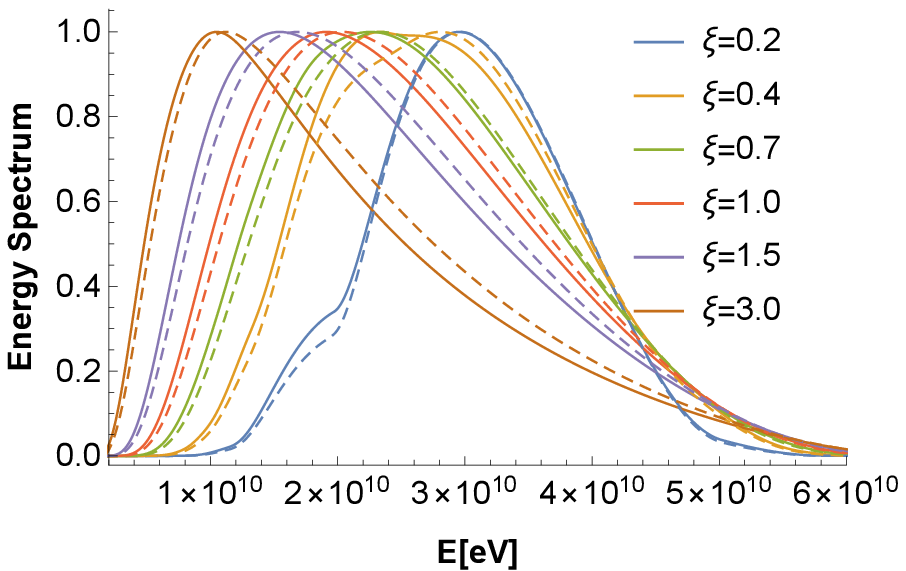}}
		\qquad
		\subfloat[][]{\includegraphics[width=0.9\linewidth]{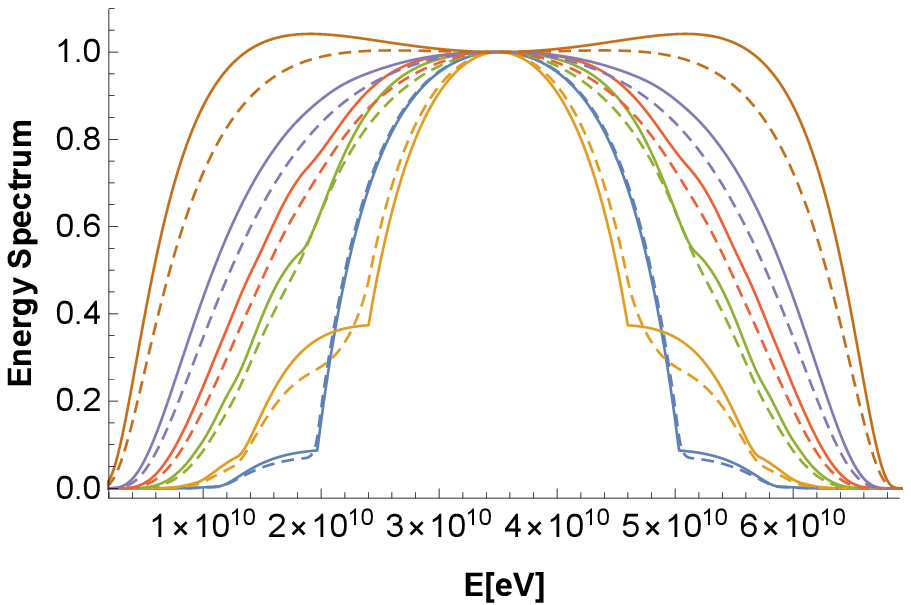}}
		\caption{Differential pair production rates $\text{d}\bar{R}/ \text{d}E_{q}$ in an infinite laser wave (solid lines) and probabilities $\text{d}\bar{W}/ \text{d}E_{p}$ for an 11-cycle cos$^2$-shaped laser pulse (dashed lines) as function of the positron energy. In panel (a) we use $\omega=1.5$eV and $E_0=16.5$ GeV for the electron beam driving the bremsstrahlung. Panel (b) refers to the second scenario with $\omega=2.4$eV and $E_0=70$GeV. Panel (c) shows the monochromatic case $\text{d}R/ \text{d}E_{q}$ and $\text{d}W/ \text{d}E_{p}$  for $E_0=70$GeV. The same color coding and normalization as in Fig.1 applies.}
	\end{figure}
	Next we consider production rates differential in the energy $E_q$ of the created positron. Note that, in our approach, the energy-differential rates with respect to the created electron are identical. Figure~2\,(a) and (b) show the bremsstrahlung-averaged rates for both scenarios. For comparison, Fig.~2\,(c) displays the monochromatic rate for $\omega'=70$\,GeV, which is discussed first. In this case, the energy distribution is always symmetric about the point $E_q\approx \omega'/2$, where it attains its maximum for moderate values of $\xi$. When $\xi$ increases, the distribution becomes broader and a symmetric pair of maxima develops to the sides, indicating again a nonperturbative above-threshold phenomenon where laser photon numbers substantially above $n_0$ provide the main contribution.
	\\
	For small $\xi$ values, the individual $n$-contributions are clearly reflected in the rates. At $\xi = 0.2$, the minimum laser photon number $n_0=2$ allows for pair creation in the energy interval between about 20\,GeV and 50\,GeV [see Eq.\,(9)] where, accordingly, step-like behaviors are observed in Fig.~3\,(c). Taking the next higher photon number $n=3$, the limits $\approx 11$\,GeV and $\approx 59$\,GeV result, which are also recognizable as steps. The steps are symmetrically located around the maximum. Their positions also represent a signature of the laser-dressed mass. Due to the intensity dependence of $m_\star$, the steps move inwards when the value of $\xi$ increases. At $\xi=0.4$, they are located at about 24\,GeV and 46\,GeV, correspondingly. We note that signatures of the laser-dressed mass have also been predicted for multiphoton pair production in other field configurations \cite{Gies} as well as for nonlinear Compton scattering \cite{Harvey,Heinzl2}.
	\\ 
	\\
	When the average over the bremsstrahlung spectrum is included,
	parts of the step-like pattern are preserved for $E_0 = 70$\,GeV
	[see Fig.~2\,(b)]. The transitions from 3 to 2 laser photons
	remain visible as kinks in the energy distributions for $\xi=0.2$ and $\xi = 0.4$. One can see, moreover, that the position of the rate maximum shifts towards lower positron energies with growing laser intensity. This is in accordance with the fact that an increasing amount of bremsstrahlung with
	lower energy plays a significant role (see Fig.~1). While the total energy of a pair satisfies the relation $E_q+E_q'\approx\omega'$, the maxima of $d\bar{R}/dE_q$ are not located at half the value of $\omega^\prime_\text{max}$ where the corresponding distribution $d\bar{R}/d\omega^\prime$ features a maximum, but somewhat below. This is because (i) each $\omega^\prime$ yields a range of different $E_q$ according to Eq.~(9), and (ii) the bremsstrahlung photons with $\omega^\prime<\omega^\prime_\text{max}$ provide larger contributions to the rate than those with $\omega^\prime>\omega^\prime_\text{max}$ (see Fig.~1). This property is particularly pronounced for the highest field intensity under consideration ($\xi=3$).
	\\
	The shifting of the rate maxima with increasing $\xi$ towards lower positron energies is also found for the LUXE parameters in Fig.~2\,(a), along with a broading of the energy distributions. When, instead, a monochromatic situation with fixed $\omega'=16.5$\,GeV is considered, symmetric bell-shaped curves result for the energy-differential rate $d\bar{R}/dE_q$ which are centered at $E_q\approx\omega^\prime/2$ (not shown). 
	\\
	It should be noted that the differential rates in Fig. 2 refer to the field-dressed energy $E_q$ of the positron---a quantity that arises naturally in an infinite laser wave model. The differential rates in Figs. 2 and 3 (see below) show that the created particles are emitted predominantly with very high energies ($E_q \approx \omega^\prime /2 \sim 10^{10}$ eV) antiparallel to the laser wave ($\theta_q \approx \pi$). As a consequence, the difference $\sim m^2 \xi^2 / \omega^\prime \lesssim 10^2$ eV between the dressed energy $E_q$ and the asymptotic free energy $E_p$ becomes very small and practically negligible in the range of $\xi$ values considered here.\\
	Figure 2 also shows the energy-differential pair production probabilities arising in a finite laser pulse. They refer to the free positron energy $E_p$, which is the appropriate quantity in a laser pulse, where the effective momentum is phase-dependent (see Sec. II. B). Taking the finite laser pulse length into account, does not change the energy spectra significantly, especially for the LUXE parameters. This is in accordance with recent studies on pair production \cite{Takabe,Kaminski} which have shown that a monochromatic description of the field can represent a very good approximation for pulse durations starting with $N\gtrsim 5$ already. Comparing Figs. 2(b) and (c), the pulse length seems to have an even smaller effect when the average over the bremsstrahlung is taken. This is understandable noting that the frequency bandwidth due to the finite duration of the pulse is of order $\Delta \omega\approx$0.1\,eV, whereas the bremsstrahlung has a spectral width of $\approx$GeV. Taking an average over this very broad spectrum thus smears out any structures much more drastically than the finite laser pulse duration does.
	\\
	\begin{figure}[htp]
		\centering
		\subfloat[][]{\includegraphics[width=0.9\linewidth]{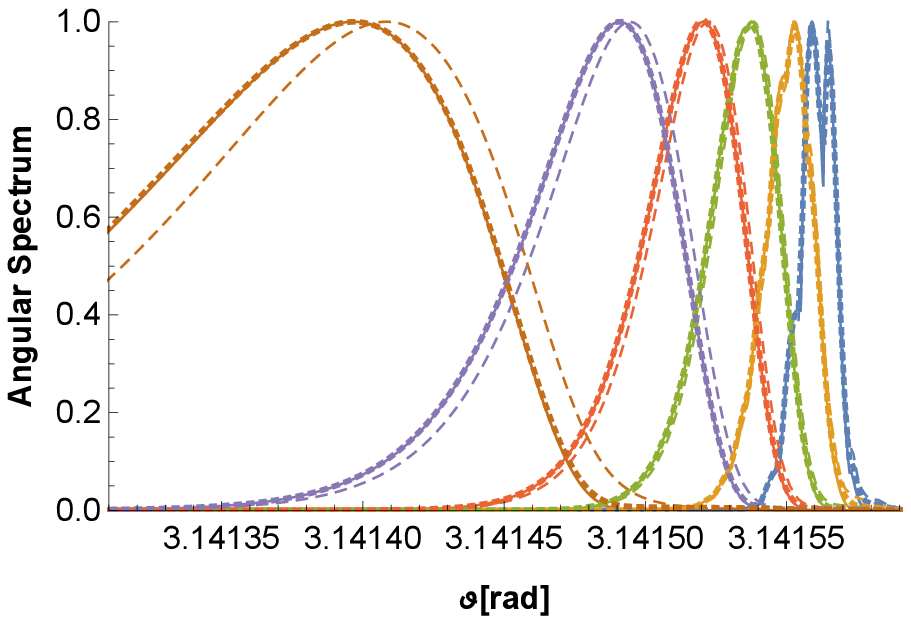}}
		\qquad
		\subfloat[][]{\includegraphics[width=0.9\linewidth]{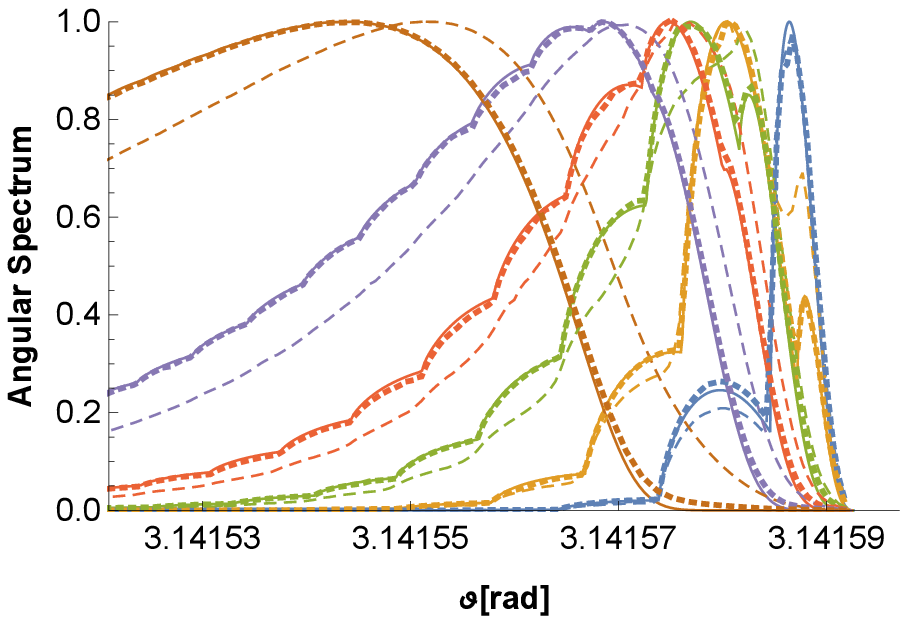}}
		\caption{Differential pair production rates $d\bar{R}/ d\vartheta_{q}$ and differential pair production probabilities $d\bar{W}/ d\vartheta_{p}$ as function of the polar emission angle of the positron. In case of the differential probability we show the LMA with $N=11$ and $f(\phi/\Phi)=\cos^2(\phi/\Phi)$ (dashed line) and also with the flat-top envelope (dotted line).  Panel (a) refers to $\omega = 1.5$ eV and $E_0=16.5$ GeV, and panel (b) to $\omega=2.4$eV and $E_0=70.0$GeV. The same color coding and normalization as in Figs.1 and 2 applies.}
	\end{figure}
	\\
	Figure~3 shows the differential production rates with respect
	to the angle $\vartheta_q$ between the positron momentum $\bf{q}$ and the laser wave vector $\bf{k}$. They possess some specific features. Overall, the created particles are emitted along angles close to $\vartheta_q = \pi$ because the energy of the bremsstrahlung photons is much larger than the energy absorbed from the counterpropagating laser wave. However, as the laser intensity increases, the transverse positron momentum grows and the width of the angular distribution is enhanced. This is in line with the shifting and broadening of the rates $d\bar{R}/d\omega^\prime$ in Fig.~1, which reflect the increasing number of laser photons that participate in the process.\\
	For the lower $\xi$ values, pronounced structures of maxima and
	minima arise in the angular distributions of Fig.~3. They can be
	attributed to the corresponding number of absorbed laser photons. To this end we note that, for given values of $\omega^\prime$ and $n$, the monochromatic rate $dR/d\vartheta_q$ exhibits a divergent maximum at $\vartheta_\text{min}$ from Eq.~(8). These singularities become finite when the bremsstrahlung average is taken. Accordingly, e.g., the peak centered at $\vartheta_q \approx 3.141587$ for $\xi = 0.2$ in Fig.~4(b) mainly results from $n=2$, with its center corresponding to $\omega^\prime \approx 62.5$\,GeV. The latter is obtained from the relation
	\begin{eqnarray}
		\omega^\prime&= &\frac{m_\star^2 n\omega(1+\cos^2\vartheta_{q})}{4n^2\omega^2-m_\star^2\sin^2\vartheta_{q}} \notag \\
		&+& \sqrt{\frac{m_\star^4n^2\omega^2(1+\cos^2\vartheta_{q})^2}{(4n^2\omega^2-m_\star^2\sin^2\vartheta_{q})^2}+\frac{m_\star^2n^2\omega^2\sin^2\vartheta_{q}}{4n^2\omega^2-m_\star^2\sin^2\vartheta_{q}}}\notag
	\end{eqnarray}
	that follows from Eq.~(6). The neighboring peak at $\vartheta_q \approx 3.1415798$ is associated with the absorption of $n=3$ laser photons. The position of the minimum between these peaks coincides with $\vartheta_\text{min} \approx 3.141585$ for $\omega^\prime= E_0$ and $n = 2$. Each minimum in the angular distribution marks another $n$ transition. Similarly as in the energy distributions of Fig.~2, the positions of the angular minima offer a possibility to observe the laser-dressed mass: due to the intensity dependence of $m_\star$, the locations of the minima are shifted when $\xi$ is changed [see Eq.~(8)], as can clearly be seen by comparing the curves for $\xi=0.2$ and $\xi=0.4$ in Fig.~3(b). The structure of angular maxima and minima is smoothed out when $\xi$ grows in the course of the perturbative-to-nonperturbative transition. For $\xi=3$ the angular distribution has become a smooth curve both in Fig.~3(a) and (b).\\
	When the finite length of a $\cos^2$-pulse is included, we see shifts towards larger angles with increasing $\xi$ (dashed lines in Fig. 3). In addition, the steps are smeared out more quickly when $\xi$ grows than for an infinite laser wave. This effect is observed to a larger extent for $E_0 = 70$GeV than for the LUXE parameters. In comparison with the energy spectra in Fig. 2, the angular distributions are thus more sensitive to the inclusion of this type of laser pulse profile. In order to reveal whether this effect is caused predominantly by the finite length or the varying intensity of the pulse, we performed additional calculations for another pulse envelope. It is composed of $\cos^2$-shaped turn-on and turn-off wings of one cycle duration each and a flat-top plateau region in between. The angular distributions resulting for this pulse profile (dotted lines in Fig. 3) lie almost completely on those of the infinite-wave calculations. Therefore, the differences found for the $\cos^2$-shaped pulse are mainly due to the intensity variations. We note besides that the energy spectra from a flat-top laser pulse coincide almost perfectly with those from the infinite-wave treatment. \\
	\begin{figure}[h]
		\centering
		\subfloat[][]{\includegraphics[width=0.37\linewidth]{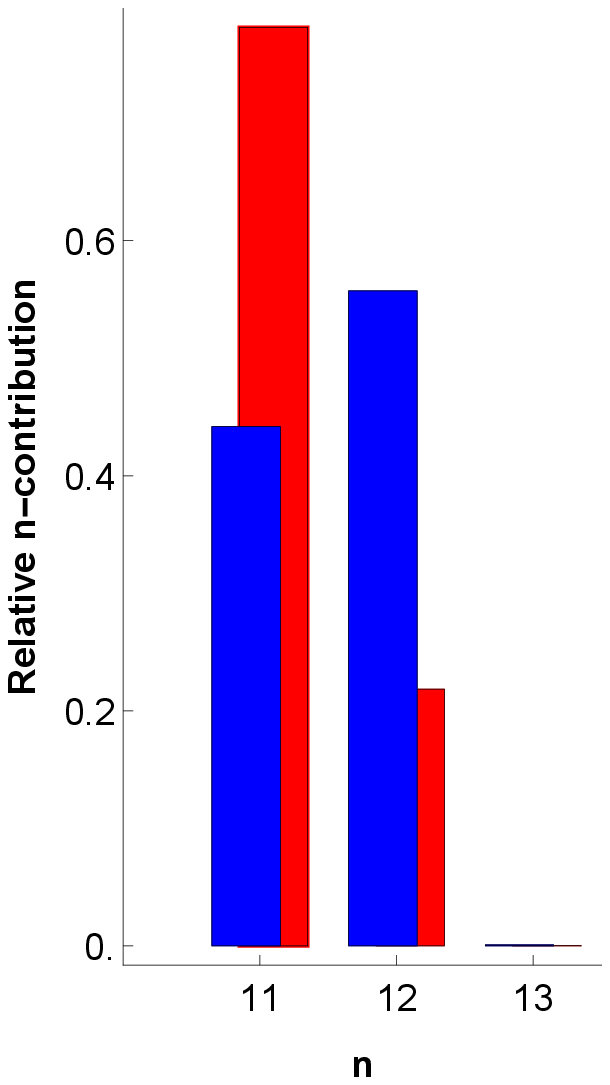}}
		\subfloat[][]{\includegraphics[width=0.63\linewidth]{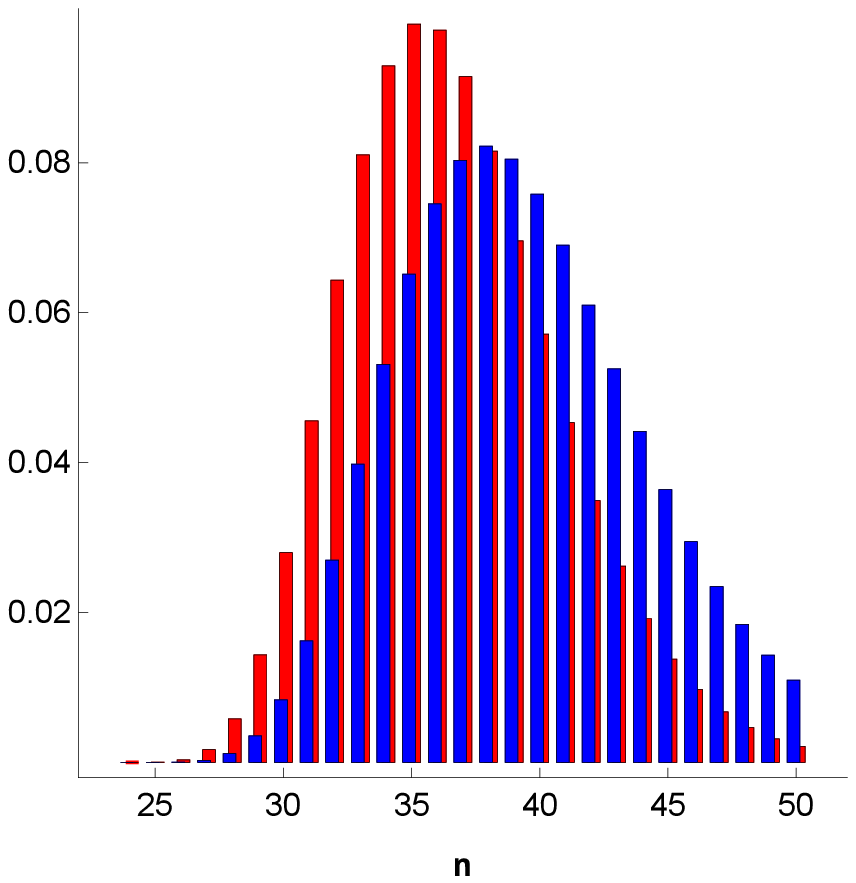}}
		\caption{ Relative contributions of individual laser photon numbers to the total pair production rate for (a) $\xi = 0.003$ and (b) $\xi = 1.0$ with $\omega = 1.5$ eV. The red bars refer to the monochromatic case $R_n/R$ for $\omega^\prime=16.5$ GeV and the blue bars to the bremsstrahlung-averaged case $\bar{R}_n/\bar{R}$ for $E_0=16.5$ GeV. For better visibility, the red and blue bars are slightly shifted against each other along the horizontal axis.}
	\end{figure}
	\\
	To analyze the impact of the finite laser pulse length and shape on the total pair yields, we note that a production probability $\bar{W}_{\rm rec} = \frac{1}{2}\bar{R} \tau$ results from the averaged rate in Eq.(13). Here, a laser pulse of duration $\tau = 2\pi N/\omega$ and rectangular shape [i.e. $f(\phi/\Phi)=1$] is assumed. For $N=11$, $E_0=16.5$GeV and, e.g., $\xi=3$ one obtains $\bar W_{\rm rec} = 1.2\times 10^{-4}$, accordingly (see Tab. I). If the pulse has instead a $\cos^2$-envelope, the production probability is reduced to $\bar{W}_{\cos} = 2.4\times 10^{-5}$. This number results from an integration over the corresponding angular distribution in Fig. 3 (or, alternatively, the energy spectrum in Fig. 2). The reduction is caused by the fact that the intensity varies over the $\cos^2$-shaped pulse and reaches its maximum value only at the center. For the flat-top pulse, the probability is $\bar{W}_{\rm flat} = 9.9\times 10^{-5}$. It amounts to a fraction of $\approx \frac{9}{11}$ of $\bar{W}_{\rm rec}$, because the plateau region spans over 9 cycles, while the turn-on and turn-off cycles give a minor contribution to the production yield only. The total pair production probabilities for all parameter sets under consideration are compiled in Table\,II.\\
	\begin{table}[h]
		\centering
		\begin{tabular}{ |p{0.4cm}||p{1.15cm}|p{1.15cm}|p{1.15cm}||p{1.15cm}|p{1.15cm}|p{1.15cm}|}
			\hline
			\parbox[c]{1cm}{ ~ \newline  ~ \newline  ~ \newline} & \multicolumn{3}{|c||}{\parbox[c]{2.5cm}{ $E_0=16.5$\,GeV \newline $\omega=1.5$\,eV}} & 
			\multicolumn{3}{|c|}{\parbox[c]{2.5cm}{$E_0=70.0$\,GeV \newline $\omega=2.4$\,eV} } \\
			\hline
			$\xi$ & $\bar{W}_\text{rec}$& $\bar{W}_\text{cos}$&$\bar{W}_\text{flat}$&$\bar{W}_\text{rec}$&$\bar{W}_\text{cos}$&$\bar{W}_\text{flat}$ \\
			\hline
			0.2  &3.8[-23]&4.2[-24]&3.1[-23] &2.1[-6]&5.6[-7]&1.8[-6]\\
			0.4  &2.3[-16]&2.7[-17]&1.9[-16]&3.6[-5]&9.7[-6]&3.1[-5]\\
			0.7  &9.7[-12]&1.3[-12]&8.1[-12]&3.7[-4]&1.0[-4]&3.2[-4]\\
			1.0  &2.3[-9] &3.4[-10]&2.0[-9]&1.3[-3]&3.8[-4]&1.2[-3]\\
			1.5  &3.3[-7]&5.5[-8]&2.8[-7]&5.0[-3]&1.5[-3]&4.3[-3]\\
			3.0  &1.2[-4]&2.4[-5]&9.9[-5]&3.0[-2]&1.0[-2]&2.6[-2] \\
			\hline
		\end{tabular}
		\caption{Total pair production probabilities $\bar{W}$ per incident electron averaged over the bremsstrahlung spectrum for different values of the intensity parameter $\xi$, with laser photon energy $\omega$ and incident electron energy $E_0$. Three different types of 11-cycle laser pulses with rectangular, $\cos^2$ or flat-top profile are considered. Note that powers of 10 are enclosed in brackets	here, i.e., $[-X] = 10^{-X}$.}
	\end{table}
	\\
	Finally, we consider the individual contributions $R_n$ and
	${\bar R}_n$ from a specific number of absorbed laser photons to the total rates $R = \sum_{n\ge n_0} R_n$ in Eq.\,(3) and, accordingly, $\bar R$ in Eq.\,(12). Figure 4 shows the ratios $R_n/R$ and ${\bar R}_n/\bar R$ for the monochromatic and bremsstrahlung-averaged cases, respectively, for the first scenario related to LUXE. In order to illustrate pair production with perturbative properties, $\xi = 0.003$ is chosen in Fig.\,4(a). Then, in the monochromatic case, the contribution from the minimal photon number $n_0 = 11$ is by far the largest; the next-to-leading order term with $n=12$ is strongly suppressed. Conversely, in the averaged case, the
	contribution from $n=12$ is larger than that from $n=11$. At first sight, this behaviour might seem to indicate already a transition from the perturbative to the non-perturbative regime, where the leading-order term is generally suppressed in comparison with higher order terms. The true reason is, however, that this redistribution effect is caused by the large interval of contributing
	bremsstrahlung energies $\omega^\prime$. (One should note that each of them has its own value of $n_0 =n_0(\omega')$.) The enhanced contribution from $n=12$ stems from somewhat lower $\omega^\prime$ values that appear with higher probability in $I_\gamma$ than those higher energies required for the $n=11$ channel. That we are still situated in the perturbative domain is confirmed by the fact that both ${\bar R}_n$ contributions scale like $\xi^{2n}$.\\
	Hence, the broad spectrum of bremsstrahlung is responsible
	for the shift of the mainly contributing laser photon numbers towards larger values. This effect persists when $\xi$ increases. For example, at $\xi=0.2$ (not shown) the channel with $n=11$--while still being energetically allowed--gives neglible contributions to both $R$ and $\bar R$ because the phase space available for the created particles is very small. Instead, the main contributions come from $n=13$, 14, 15 and 16 amounting respectively to about $35\%$ ($22\%$), $41\%$ ($42\%$), $17\%$ ($25\%$) and $4\%$ ($8\%$) for the monochromatic rate $R$ (averaged rate $\bar R$). Here, the pair creation already exhibits nonperturbative features. Entering more deeply into the nonperturbative regime at $\xi = 1.0$, we see in Fig.\,4(b) a pronounced shift towards larger laser photon numbers due to the averaging over the bremsstrahlung.\\
	\\
	Before moving on to the conclusion, we point out that the special features found in our second scenario (see Figs. 2 and 3) might be accessible to the experimental initiative of LUXE as well. By irradiating a solid target foil with the envisaged optical laser beam that is going to reach maximum intensities above $10^{20}$ W/cm$^2$ [14], high harmonic radiation could be efficiently generated from the laser-driven plasma surface \cite{HHG}. If the 7th harmonic was extracted and collided with the bremsstrahlung beam, then the product $\omega E_0$ (where $\omega$ now refers to the frequency of the 7th harmonic) attains practically the same value as in the second scenario--this way opening prospects to detect the pronounced few-photon effects and laser-dressed mass signatures occuring there. By exploiting even higher harmonics, also the linear Breit-Wheeler process might become observable this way.
	\section{Conclusion}
	Strong-field Breit-Wheeler pair creation with bremsstrahlung
	gamma-photons in the perturbative-to-nonperturbative transition regime has been analyzed. We have shown that the relevant spectral range of bremsstrahlung, that mainly contributes to the production rate, shifts to smaller frequences the larger the laser intensity parameter $\xi$ is. A similar shifting has been found, accordingly, in the energy spectra of created particles that, in addition, become more and more asymmetric. For the number of absorbed laser photons an opposite shifting towards higher values occurs. The broad bremsstrahlung spectrum also influences the angular distributions of created particles in a characteristic manner. Photon-number transitions $n\to n+1$ are clearly visible therein, whose positions are linked to the laser-dressed mass. For moderate values of $\xi$, signatures of $m_\star$ can also be identified in the particles' energy spectra. The finite laser pulse length and shape was generally found to be of minor relevance in the considered parameter regimes. Only the angular distributions show a more pronounced sensitivity to the precise shape of the applied field pulse. 
	\\
	\begin{acknowledgements}
		This work has been funded by the Deutsche Forschungsgemeinschaft (DFG) under Grant No. 416699545 within the Research Unit FOR 2783/1.
	\end{acknowledgements}

\end{document}